# Local shear induces long-range suppression of cellular motion in stiff epithelial monolayers

Shahar Nahum[1], Adi Y. Elkabetz[1], Matan Elbaz[1], Liav Daraf[1], Yael Lavi[1], Lior Atia[1]*

[1]Department of Mechanical Engineering; Ben-Gurion University of the Negev, Beer-Sheva 8410501, Israel; *atialior@bgu.ac.il

## Author contributions

S.N and L.A. conceptualized the research. S.N., Y.L., and L.A. designed the experiments. S.N., A.E., and Y.L. performed the experiments. M.E., and L.A. designed and created the experimental system. S.N., A.E., and L.A. designed and performed all presented analyses. S.N., Y.L., L.D., and L.A. contributed to data interpretation. S.N and L.A. wrote the manuscript. L.A oversaw the project.

## Competing interest statement

Authors declare that they have no competing interests.

## This PDF file includes:

Main Text
Figures 1 to 4
Supplementary Figures 1 to 9
Supplementary Table 1
Supplementary Videos 1 to 2 (referrals)

## Abstract

As small particles skim our airways during breathing, or our intestines during digestion, the surface epithelium is subjected to local exogenous shear that deforms hundreds to thousands of tightly interacting cells. Unlike shear deformations applied at the macro-tissue scale or the micro-cell scale, the effects of such perturbations at the meso-scale remain largely unexplored. To address this, we developed a mesoscopic probe that adheres to the apical surface of an epithelial monolayer and applies magnetically driven local shear. We find that localized shear suppressed cellular migratory dynamics far beyond the immediate neighbors in stiffer layers, whereas in softer layers the same perturbation produced no detectable long-range change in dynamics. This mechano-transductive relationship is further supported by unconfined mature layers, in which increased stiffness was accompanied by restored shear responsiveness. Viewed at the level of collective dynamics, shear-induced migratory suppression in stiff layers was often accompanied by reduced MSD scaling exponents and changes in cell shape, but these responses were not fully captured by the epithelial jamming framework. Together, these observations provide a new perspective on how a local mechanical perturbation traverses the epithelial monolayer to influence both nearby and distant cellular environments.

## Significance

Epithelial tissues are routinely exposed to localized mechanical shear, arising from particles or perturbing objects. While the effects of shear forces acting at single-cell or whole-tissue scales are well-studied, how localized perturbations influence collective epithelial behavior is unclear. Here, we show that local shear suppresses cellular motion in a manner that largely depends on epithelial stiffness. In mechanically stiff epithelia, localized shear induces long-range suppression of cellular dynamics across many cell rows, whereas in softer epithelia, the same perturbation produces no detectable long-range suppression. These findings point to epithelial stiffness as a potential mechanical filter that shapes the spatial reach of local shear perturbations, such as those arising from an inhaled particle or an emerging tumor, and thereby their capacity to reorganize collective behavior.

## Main text

## Introduction

Epithelial layers line organs, maintain barrier function, and actively remodel during development, homeostasis, and repair. While doing so, they experience mechanical shear forces acting along multiple size scales. Large tissue-level shear deformations that occur during morphogenesis, breathing, or peristalsis were studied using various whole-tissue stretching assays, showing fundamental growth-related pathways[1-3], and mechanical fracturing characteristics[4,5]. Small cell-level shear deformations that occur during immune extravasation[6], mitosis[7,8], or extrusion[9], were studied using subcellular perturbations and were shown to affect cell adhesion[10], cytoskeleton organization[11], and even promote intracellular fluid-like microrheology[12-14]. However, between these regimes lies an unexplored mesoscale domain, in which, for example, exogenous particles in the lung or in the intestines apply locally concentrated shear forces that deform hundreds to a few thousand of tightly interacting epithelial cells. Mesoscale forces have been studied in the developing mesoderm[15,16], however, how such mesoscale shear extends across the adherent epithelial monolayer, and how it contributes to local mechanotransduction in relevant physiological contexts, remains largely unknown[17].

To fill this gap, we developed a mesoscopic magnetic probe that adheres to the epithelial apical surface and applies controlled lateral shear to living monolayers. By characterizing local cellular dynamics and rheological properties, we demonstrate that the shear response extends far beyond the perturbing site in an unexpected manner, and that it depends strongly on local tissue rigidity. These results identify mechanical rigidity as a key parameter governing the spatial reach of shear-induced signals and establish localized shear as a distinct mode of multicellular communication.

## Local mechanical shear suppresses cellular motion in nearby cells

To examine how local mesoscale perturbations affect epithelial behavior, we established an assay that applies controlled mechanical shear to confined epithelial monolayers. The system consists of a maturing tissue of Madin-Darby Canine Kidney (MDCK) epithelial cells, cultured within physical confinement and allowed to mature for three distinct durations: right as the layer becomes confluent (t0); 5 hours post-confluence (t5); and 10 hours post-confluence (t10). We used a mixture of polydimethylsiloxane and iron oxide (PDMS-Fe) to create a magnetizable disc. The disc was placed directly atop the monolayer (Fig. 1a,c; Methods; Supplementary Fig. 1), and a magnetic tweezer setup was used to apply a lateral force of approximately $2 \times 10^4$ nN (Fig. 1b; Methods; Supplementary Fig. 2), generating a 10s duration localized shear (Supplementary Video 1; Methods). As a control, we prepared matched samples with attached magnetic discs but without applying any force. Monolayers were then imaged using time-lapse to capture cells migratory response. This design enabled us to examine how maturation and localized shear influence the mechanical and dynamic properties of epithelial tissues surrounding the perturbed site. Across individual experiments, mean cellular speed showed no consistent systematic change over the imaging period (Supplementary Fig. 3), supporting its averaging over time for the comparisons to follow. Surprisingly, unlike studies applying tissue-scale deformations[18,19], our localized shear caused motility to fall well below that of unperturbed tissues, marking a pronounced suppression of migration (Supplementary Video 2). That migratory suppression was observed far beyond the disc in the t0 and t5 tissues (Fig. 1d-e;

Methods). In contrast, the more mature t10 tissues showed no comparable reduction in motion, as mean speed remained unchanged between sheared and control samples. An ROI-label permutation test using a KS-distance statistic further showed shear-associated shifts toward lower speeds at t0 and t5 ($p = 0.0383$ and $p = 0.0252$, respectively), but not at t10 ($p = 0.9007$) (Methods, Supplementary Fig. 8). To examine the spatial and temporal development of this response to shear, we calculated speed profiles across the horizontal extent of the ROI at selected time points and constructed corresponding space–time kymographs (Supplementary Fig. 9). In the responsive conditions, the reduction in speed was already distributed across the full 402μm ROI at the first recorded time-lapse point, and no subsequent proximal-to-distal progression was resolved during imaging. Because the mechanical perturbation was applied locally at the magnetic disc, the response observed throughout the distant ROI indicates that the shear-induced signal had propagated across at least the full imaged field before the first time-lapse imaging. Thus, the propagation occurred faster than could be resolved by the present imaging protocol, and its kinetics could not be characterized directly. This observed relationship between tissue maturation and its dynamic responsiveness to shear prompted us to consider whether this is due to progressive mechanical stiffening.

## Maturation-dependent softening limits long-range shear-induced responses

To explore the idea that maturation alters tissue mechanical responsiveness, we quantified the shear strain, $\gamma(t) = d(t)/h$, generated within the epithelial layer directly beneath the magnetic disc, where $d(t)$ is the lateral displacement of the disc and $h$ is the mean tissue height for the corresponding maturation condition (Fig. 2a–b; Methods; Supplementary Fig. 4). Tissue height was measured in paraformaldehyde-fixed samples (Methods). The measured strains were substantial, reaching values of order 0.1–1, and were therefore large enough to potentially contribute to strain-induced cytoskeletal fluidization.[20] The strong motility suppression in t0 and t5, and the absence of this response in t10, initially led us to hypothesize that the less mature tissue was compliant and softer, whereas the more mature tissue was incompliant and stiffer. Unexpectedly, the strain measurements revealed the opposite. The less mature tissues showed substantially lower strain than the mature tissues, identifying the most mature ones as the softest (Fig. 2c; Supplementary Table 1). This result indicates that, under the prescribed spatial constraints (Fig. 1a), maturation leads to a tissue softening transition. Notably, since t0 and t5 experience comparable strains, that transition is not progressive but rather abrupt (Fig. 2c).

Power-law rheology is commonly used to characterize the time-dependent deformation of biological cells and tissues. We fitted the strain trajectories as $\gamma(t) = A(t/t_{\mathrm{ref}})^{\beta}$, where $A$ describes the magnitude of deformation at the reference time of 1s and $\beta$ describes its time dependence. Note that the exponent $\beta$ is not itself a measure of stiffness. Rather, smaller values indicate more solid-like behavior, whereas larger values indicate more fluid-like behavior and, for a given $A$, greater accumulation of deformation over time. We therefore examined whether the maturation-dependent softening was accompanied by changes in $\beta$. Fitting the strain curves (Supplementary Fig. 5) to the power-law model yielded two key findings. First, during monolayer recovery after shearing, $\beta$ values markedly decreased, but in a manner that was independent of tissue maturation (Fig. 2d, Supplementary Fig. 6a, b). This behavior is consistent with previous reports showing that external forces transiently fluidize the cytoskeleton[12]. Because cytoskeletal remodeling can depend on strain rate,[20] we also examined whether the relative change in $\beta$ between the loading and release phases was associated with the initial strain rate (Methods). We found no consistent relationship between the initial strain rate and the relative change in $\beta$ across the experimental conditions (Supplementary Fig. 6c). Second, $\beta$ remained essentially unchanged throughout all maturation times (Fig. 2d). Thus, the maturation-dependent softening we observe is not captured by the exponent $\beta$.

The overall effects of local shear deformations act to suppress cellular dynamics many cell rows away from the perturbed site in the less mature tissue (Fig. 1d; Supplementary Fig. 9; Supplementary Video 2). Yet, this less mature tissue also undergoes small shear-induced deformations. Taken together, the observations lead to a striking paradox. If the extent of deformation were the primary determinant of the long-range suppression of cellular dynamics, one would expect the most strained tissue (t10; Fig. 2c) to show the strongest long-range effect. Instead, we observe the opposite: in the less mature tissues (t0 and t5), motility suppression extends across the distant imaged field, while in the most mature tissues (t10), no comparable long-range suppression is detected (Fig. 1d; Supplementary Fig. 9). This counterintuitive behavior raises the possibility of a distinct physical mechanism. Analogous to pressure waves traveling farther in more rigid materials, could shear-induced signals have a greater spatial reach across a stiffer epithelial monolayer?

## Induced migration promotes shear responsiveness

Further exploring the relationship between tissue stiffness and shear responsiveness might require a more detailed mapping of spatial stiffness. We lacked, however, the technology to precisely control the micro-level placement of the magnetic disc. Nevertheless, we were able to gain an initial clue by plotting the stiffness measurements obtained from randomly placed discs against their distance from the tissue edge (Fig. 3a). We found a modest tendency for stiffness to gradually decrease with increasing distance from the tissue boundary. This observation made us consider a possible connection to the well-known phenomenon in which migration velocity is generally higher near the free tissue boundary[21] (Fig. 3b). It led us to hypothesize that enhancing cellular migration within the t10 tissue could act to stiffen it and thereby restore its susceptibility to shear-induced migratory suppression.

We therefore performed a complementary set of experiments in which, immediately after confluency, we removed the physical barrier confining the monolayer in order to promote cellular migration (Fig. 3c; Methods). We then waited either 5h or 10h, applied the same shear protocol as before to these unconfined tissues, and measured the rheological response. The resulting strain curves followed a clear power-law, yielding $\beta$ values that closely matched those of confined tissues (Fig. 3d). Furthermore, similar to the confined monolayers, during recovery after shearing of the unconfined monolayers, the $\beta$ values consistently dropped (Fig. 3d). That is, no behavioral changes were captured by the exponent $\beta$ for unconfined monolayers. However, shear strains in the unconfined monolayers were substantially lower than those in their confined counterparts (Fig. 3e). This reduction supported our hypothesis and demonstrated that unconstrained maturation promotes a mechanically stiffer phenotype. While confined t10 tissues showed no detectable change in mean cellular motility following shear (Fig. 1), unconfined t10 tissues exhibited a pronounced reduction in cellular speed following shear (Fig. 3f-g). Consistently, the ROI-label permutation test showed shear-associated shifts toward lower speeds at both t5UC and t10UC ($p = 1.60 \times 10^{-4}$ and $p = 0.0295$, respectively) (Methods, Supplementary Fig. 8). Together, these findings demonstrate that lifting spatial constraints increases stiffness and restores long-range shear-induced motility suppression, strengthening the idea that mechanical cues transmit more effectively in stiffer monolayers.

Previous studies have shown that processes such as wound healing or stimulated migration can trigger cell-cycle reentry and increased proliferation[19,22]. We therefore asked whether the enhanced dynamics and applied shear in unconfined tissues might have induced cell division, potentially contributing to the observed mechanical changes (Fig. 3). Using a cell cycle fluorescence reporter[19,23], we quantified cell-cycle states in both perturbed and unperturbed conditions[24](Methods). We detected no measurable differences in cell-cycle progression, indicating that the mechanical perturbation did not affect proliferation and is therefore unlikely to account for the observed behaviors in unconfined tissues (Supplementary Fig. 7).

## Shear-induced migratory suppression partially coincides with jamming metrics

To gain a deeper physical understanding of our observations, we turned to a theoretical framework inspired by concepts from granular and soft glassy systems[25], known as multicellular jamming. This framework has emerged in biophysics as a way to elucidate observed epithelial transitions from an unjammed fluid-like state with abundant cellular motion, to a jammed solid-like state with suppressed motion[26-38]. Since we also observed a shear-induced suppression of cellular motion, we wondered whether our observations are consistent with, and can be described by, dynamical and geometrical signatures identified with epithelial jamming.

Dynamically, the mean squared displacement (MSD) quantifies how far cells wander from their original positions over $\Delta t$ time (Fig. 4a). MSD distinguishes between random diffusion, confined motion, and super-diffusive rearrangements[33,37,38]. By fitting the MSD curves to a power-law model ($MSD(\Delta t) \propto t^{\alpha}$) we could extract the change in $\alpha$ parameter that reflects altered caging dynamics. By definition, $\alpha > 1$ reflects an uncaged dynamics, and reducing $\alpha$ values reflects an increasing caging (Fig. 4a). In most cases, applying local shear shifted the cell-specific $\alpha$ values toward lower values (Fig. 4b). Mann–Whitney comparisons supported this shift at t0 ($p = 0.0143$) and t5UC ($p = 2.23 \times 10^{-26}$). At t5, the comparison yielded $p = 0.0527$, only slightly above the conventional 0.05 threshold, with the median changing in the same direction. No comparable change was observed at t10 or t10UC. Thus, changes in $\alpha$ broadly, but not uniformly, paralleled the changes in cell speed (Fig. 3f). Most notably, t10UC showed reduced speed without a corresponding change in $\alpha$. Nevertheless, in general, the observations suggest that shear reduces the magnitude of cellular self-propulsion and that the collective pattern of motion transitions toward more caged dynamics. We then wondered whether we could support or refute this interpretation with other jamming metrics, namely geometrical.

Geometrically, a central metric in the jamming framework is cell shape, as formalized in the theoretical vertex model[27,28,39]. In this model, multicellular jamming emerge from the balance between cell–cell adhesion and cortical tension, which are encoded in cell shape[27]. Multiple studies corroborated the main prediction of the vertex model, showing that a transition from an unjammed state to a jammed state is reflected by a reduction in cell shape indices ($Pereimeter/\sqrt{Area}$ ; Aspect Ratio)[29,33-35,37,38]. In most of our observations, applying local shear reduced the cell shape values (Fig. 4c). However, in the unconfined t10 tissue, in which $\alpha$ did not change, but the speed decreased, cell shape increased. This important exception might be reflecting missing unknown factors in the jamming framework.

## Discussion

The results presented here establish a direct link between epithelial rigidity and the spatial reach of mechanically induced suppression of cellular dynamics. Local apical shear, when applied to a mechanically stiff epithelial layer (t0, t5, t5UC, t10UC; Fig. 2c and Fig. 3e), induces suppression of migration across many cell rows, far beyond the site of perturbation. In contrast, in the softer t10 layers, the same shear perturbation did not produce detectable long-range suppression of cellular dynamics (Fig. 2c). These observations identify the effective mechanical stiffness of the monolayer as a predictor of the spatial reach of shear-induced response. Because epithelial monolayers are active and prestressed, this effective stiffness may reflect, at least in part, actomyosin-generated contractile tension.

In an actively prestressed epithelial layer, greater contractile tension can increase resistance to additional deformation, thereby contributing to the lower strains measured under a fixed applied force. Previous measurements in airway epithelial layers showed that unjammed tissues exhibited significantly greater intercellular tension and a tendency toward greater substrate tractions, while the spatial correlations of tension decayed more rapidly.[33] Thus, the magnitude of contractile tension and its spatial coherence need not vary together. In our system, the long-range response observed in the mechanically stiffer layers may therefore reflect differences in actomyosin-generated prestress, in the spatial coordination of contractile forces, or in both. Because traction forces and intercellular stresses were not measured in our rigid-substrate configuration, these possibilities cannot presently be distinguished.

Long-range mechanical coupling between epithelial cells can also be mediated through deformation of compliant substrates, as previously demonstrated for MDCK monolayers cultured on polyacrylamide gels.[40] In the present study, however, the monolayers were cultured directly on rigid tissue-culture dishes rather than on compliant gels, specifically to minimize deformation of the underlying substrate during magnetic-disc pulling. Therefore, the observed long-range suppression is unlikely to result from strain transmission through a deformable substrate. This configuration does not exclude a role for cell–substrate adhesions in transmitting forces within the monolayer, and determining how substrate compliance modifies the response to localized shear remains an important question.

Related studies provide complementary rheological context for our assay. Autonomously migrating MDCK monolayers flowing around a fixed circular obstacle were shown to behave as a Maxwell viscoelastic liquid, with tissue deformation relaxed through cell rearrangements.[41] Simulations using a continuously forced, cell-sized probe further showed that local yielding and compliance depend on both cell motility and cellular deformability.[42] Similarly, simulations of a force-driven disc in an epithelial-like sheet showed that, without active cell motility, the disc stalls against the passive elastic response, whereas active cellular remodeling enables sustained probe motion and lowers the effective tissue viscosity.[43] Although these steady-flow and continuously driven configurations differ from our brief apical-shear assay, together they support the broader view that the response to a localized mechanical perturbation is governed jointly by tissue rheology and active cellular dynamics.[41-43]

All examined tissues exhibited relatively low MSD exponents, while their mean shape indices remained above the canonical vertex-model threshold of approximately 3.81.[33,44] We therefore do not interpret shear as driving the layers across a jamming transition. Nevertheless, in most conditions, local shear reduced cell-shape metrics and cellular motion in parallel. Consistent with this distinction, previous studies have shown that epithelial geometry and dynamics can vary progressively along jamming-associated trajectories, with decreasing cell elongation generally accompanying slower and more constrained dynamics.[44-46] In this context, the concordant directional reductions in speed, MSD exponent, and cell shape observed at t0, t5, and t5UC are consistent with a shift toward more constrained dynamics, whereas the divergent responses at t10 and t10UC indicate that these metrics are not universally coupled.

A unifying interpretation of our data is that shear suppresses residual cellular dynamics within an already mechanically constrained system. In this view, self-propulsion prior to shear serves as an active perturbation that sustains residual cellular motion and shape irregularity, keeping cells mechanically poised and responsive. Local shear suppresses these dynamics, shifting the tissue toward reduced motion and, in most conditions, greater geometric regularity. Because epithelial unjamming and collective migration have been associated with increased glycolytic activity and metabolic demand,[31] our findings raise the possibility that the shear-induced suppression of residual dynamics is accompanied by a reduction in metabolic demand. However, metabolic state was not measured here, and previous work has shown that metabolic and migratory changes can be spatially decoupled.[31] Whether the mechanical response observed here has a metabolic correlate, and whether such a change precedes or follows the suppression of motion, remains an open question.

Taken together, these findings demonstrate that reduced motility, more constrained dynamics, and increased mechanical stiffness do not necessarily covary in epithelial monolayers. Hence, the intuition drawn from passive granular and glassy materials,[25] according to which more constrained or more jammed-like behavior necessarily implies greater stiffness, does not apply straightforwardly to this active epithelial system.

# Methods

## Cell Culture and Experimental Groups

Madin-Darby Canine Kidney (MDCK) II epithelial cells (stably expressing either ZO-1 GFP[47], or the FUCCI transgene[19]) were used throughout this study. FUCCI Cells were a kind gift from Lars Hufnagel. Cells were maintained in Modified Eagle Medium (MEM) supplemented with 10% fetal bovine serum (FBS), 1% penicillin–streptomycin and 1% L-Glutamine, under standard culture conditions of 37°C, 5% $CO_2$, and 85% humidity. For each experiment, a 35 mm rigid tissue-culture dish (Nunclon Delta, ThermoFisher; polystyrene - Young Modulus of $\approx 3\,Gpa$) was prepared by attaching a single Ibidi insert (Culture-Insert 2 Well cat#80209) near the dish's edge to allow proper access for the magnetic tweezer. The insert was either new or prewashed with ethanol prior to use. One of the wells was seeded with 56,000 cells in 70 µL. Cells were cultured directly on the rigid dish surface; no compliant polyacrylamide gel or other deformable substrate was used. This configuration was selected to minimize deformation of the underlying substrate during magnetic-disc pulling. To prevent dehydration, 2 mL of additional medium was added around the well. Cells were cultured for defined durations prior to shear: t0 - 20 hours to reach confluency; t5 - 25 hours; t10- 30 hours. Two experimental groups were defined based on when the insert was removed: Unconfined (UC) group: The well was removed after 20 hours, allowing the tissue to expand freely; Confined group: Cells remained confined within the well throughout the growth period.

## Preparation of Magnetic Discs

Magnetic discs used for pulling experiments were fabricated in-house using polydimethylsiloxane (PDMS; SYLGARD™ 184 Silicone Elastomer KIT, DOW) mixed with 10% (weight) iron oxide (Iron(II,III) oxide $Fe_3O_4$, Sigma-Aldrich, Cat# 637106) to allow magnetization. A PDMS mixture with the iron and 10% curing agent was prepared. The mixture was thoroughly mixed, degassed under vacuum for 10 minutes, and cast on ethanol-cleaned glass slides to a final thickness of ~500 µm. The thickness control was achieved by metallic spacers (filler gages; Amazon). After curing at 150°C for 10 minutes and cooling in room temperature, the PDMS was removed from the glass and cut into 750 µm diameter discs using a biopsy punch (Reuse Biopsy Punch, 0.75 mm, Cat# 504529, WPI).

Following fabrication, the PDMS-Fe discs underwent a multi-step surface treatment protocol to enhance adhesion to the epithelial tissue. Discs were first immersed in 70% ethanol and exposed to UV light for 30 minutes to sterilize and activate the surface. Next, the discs were washed three times with 3 mL of PBS. An Eppendorf tube containing the 20 µL SANPAH (Sulfo-SANPAH (sulfosuccinimidyl 6-(4′-azido-2′-nitrophenylamino) hexanoate, Moshe Shtauber) solution was thawed, then diluted with 980 µL of PBS and added to the beaker with the discs. The beaker was then placed under UV light for an additional 30 minutes to activate the crosslinking of SANPAH to the PDMS surface. After UV activation, the discs were washed with 3 mL PBS as many times as was necessary to render the beaker completely clean of SANPAH (the solution is completely clear).

For RGD coating, a 50 µg/mL RGD peptide (Arg-Gly-Asp, Sigma-Aldrich, Cat# A8052) stock solution in PBS was prepared. Of this solution, 200 µL were diluted with 3.8 mL PBS, and added to the beaker with the discs. Finally, the beaker with the discs and RGD solution was fixed to a shaker using aluminum tape and shaken at 370 RPM for 24 to 72 hours at room temperature.

## Disc Placement and its Adhesion to the tissue

To allow for discs adherence to tissue, a permanent round magnet (Amazon 1.94 mm diameter, 0.96 mm thick) was taped to the underside of the dish at the desired location. Using tweezers, a disc was gently introduced into the medium above the magnet, allowing it to sink and adhere to the tissue, such that it was attracted solely by the magnet. After disc placement, samples were incubated for 30 minutes at 37°C, 5% $CO_2$, and 85% humidity to promote strong disc–tissue adhesion.

## Local shearing

To apply local shear forces, the culture well and the magnet were carefully removed. The sharp tip of the magnetic tweezer was positioned 500 µm from the disc and connected to a current supply set to 2A. 10 second duration force was applied on the disc, followed by 10 seconds recovery, producing 20-second videos that captured both deformation and relaxation phases of the tissue response (Supplementary video 1).

## Microscopy

## Long-term imaging (cellular migration)

To examine long-term responses of the tissue to local shear, time-lapse imaging was conducted following the shearing procedure. Real-time imaging was carried out on a ZEISS AXIO OBSERVER 7 microscope, equipped with an environmental control chamber (maintained at 37°C, 5% $CO_2$, and 85% humidity), and an AXIOCAM 712 MONO camera. Samples were placed under incubation-controlled conditions and imaged every 10 minutes for 3 to 5 hours.

## Short-term imaging (mesoscale rheology)

Short-term imaging was performed using a ZEISS AXIO VERT.A1 microscope equipped a AXIOCAM 202 MONO camera. Shear experiments were imaged in real time at 31 frames per second. Imaging was performed in both phase contrast and fluorescence modes, allowing precise observation of cell boundaries and tissue morphology.

## Optical Sectioning imaging

Tissues were fixed and stained for actin (alexa 647-phalloidin). Z-stacks were obtained using ZEISS AXIO OBSERVER 7 microscope equipped with an Apotome – a structured illumination module - to obtain optical sectioning of the tissue.

## Tissue Height Measurements

Z-stacks were processed using the 3D Viewer tool in Fiji, allowing visualization of the actin-labeled cell architecture along the apical–basal axis. To quantify tissue height, four representative slices were selected from each sample, and the vertical extent of individual cells was measured manually using the PlotDigitizer (free version) software.

## Image Velocimetry

To analyze tissue deformation and flow fields, we used TPIV (Time-resolved Particle Image Velocimetry), a Fiji plugin[48]. PIV was performed on phase contrast images to calculate motion vector fields, including direction and magnitude of local movements. Data included vector length, direction (angle), and terminal positions, allowing speed measurements throughout the experiment. Velocity fields were calculated using interrogation windows of 32 × 32 pixels with 50% overlap. Vector estimation was refined using a sub-pixel accuracy parameter of 10, and spurious vectors were filtered using the median-correction algorithm implemented in TPIV.

## Actin staining

To visualize actin structures and measure epithelial tissue height, samples were fixed with 4% paraformaldehyde (Formaldehyde Solution, 16% (w/v), Thermo Scientific, Cat# 28908 diluted one ampule in 30ml PBS), permeabilized with 0.1% Triton X-100 (Triton X-100, 1% stock solution, Rhenium, Cat# HFH10Triton or X-100, 10% stock solution, Rhenium, Cat# TS-28314) and blocked using 1% BSA (Bovine Serum Albumin, Reagent Grade, Millipore, Cat# 81-066-4) in PBS. F-actin was then stained with Alexa Fluor 647-conjugated phalloidin (Rhenium, A22287), incubated for 60 minutes at room temperature in darkness, and subsequently washed with PBS. All samples were stored in PBS at 4°C, protected from light, and imaged within 1 to 2 weeks.

## Segmentation and Tracking

Segmentation of the discs was performed using the Pixel Classification workflow in Ilastik (v3.4.3). The classifier was trained on several representative frames to distinguish between disc and cellular regions. The resulting probability map was thresholded in ImageJ using the "Make Binary" tool with auto-thresholding. A custom MATLAB script computed the disc displacement vector, assuming unidirectional movement.

For cell segmentation, Cellpose was employed to detect individual cells. The Cyto 2 model was used for phase contrast images. This enabled detailed analysis of morphology, including area, aspect ratio, and circularity. Cell tracking was performed using TrackMate (Fiji plugin), which is used for tracking labeled ROIs across frames, producing positional and morphological data for each cell.

## Shear strain and strain-rate analysis

The shear strain was calculated as $\gamma(t) = d(t)/h$, where $d(t)$ is the magnetic-disc displacement and $h$ is the mean tissue height measured for the corresponding experimental condition. The strain rate was calculated numerically from the shear-strain trajectory using the MATLAB *gradient* function. It uses centered finite differences at interior time points and one-sided differences at the endpoints. The relative change in the power-law exponent between the pull and release phases was calculated for each independent disc pull as

$$\Delta\beta_{\text{rel}} = \frac{\beta_{\text{pull}} - \beta_{\text{release}}}{\beta_{\text{pull}}}.$$

The relationship between the initial strain rate and $\Delta\beta_{\text{rel}}$ was assessed separately for each experimental condition using Pearson correlation analysis, as shown in Supplementary Fig. 6c.

## Statistical Analysis

**ROI-level comparison of PIV-speed distributions.** Because PIV grid values within the same FOV are spatially and temporally correlated, the statistical analysis was designed not to depend on the number of PIV grid points. Each ROI corresponded to an originally acquired, non-overlapping FOV and was treated as one spatial analysis unit contributing one complete PIV-speed distribution. The null hypothesis was that the sheared and non-sheared ROIs were sampled from the same underlying distribution of PIV speeds. Under this hypothesis, the shear labels assigned to the ROIs are exchangeable. We therefore compared the two groups using an ROI-label permutation test with a Kolmogorov–Smirnov-distance statistic. For each ROI, an empirical cumulative distribution function (CDF) was calculated from all retained PIV-speed values across the spatial grid and analyzed frames. The ROI-level CDFs were then averaged separately within the sheared and non-sheared groups on a common speed axis, giving equal weight to each ROI (Supplementary Fig. 8, left). The observed statistic, $D_{\mathrm{obs}}$, was defined as the maximum absolute difference between the mean sheared and non-sheared CDFs. To generate the null distribution, the ROI labels were reassigned while preserving the original numbers of sheared and non-sheared ROIs, and the group-mean CDFs and corresponding KS distance, $D_{\mathrm{perm}}$, were recalculated for each assignment. Three non-overlapping ROIs were analyzed per experiment; adjacent ROIs were separated by approximately 10 $\mu$m, and PIV analysis was performed separately for each ROI. This procedure follows established principles of nonparametric permutation inference for spatially structured imaging data, in which measurements within each analysis unit are retained while inference is based on reassignment of exchangeable higher-level units rather than on the number of pixels or grid points.[49,50] All possible ROI-label assignments were evaluated for t0 $\left[\binom{18}{9} = 48{,}620\right]$, t5 $\left[\binom{15}{6} = 5{,}005\right]$, and t10 $\left[\binom{15}{6} = 5{,}005\right]$. For t5UC $\left[\binom{27}{12} = 17{,}383{,}860\right]$ and t10UC $\left[\binom{30}{15} = 155{,}117{,}520\right]$, a Monte Carlo approach was used, in which 100,000 ROI-label assignments were sampled uniformly with replacement from the complete set of possible assignments. For t0, t5, and t10, for which all possible assignments were evaluated, the exact $p$-value was calculated as the fraction of assignments satisfying $D_{\mathrm{perm}} \geq D_{\mathrm{obs}}$ (Supplementary Fig. 8, right). For the Monte Carlo analyses, the $p$-value was calculated as $(b + 1)/(B + 1)$, where $b$ is the number of sampled assignments satisfying $D_{\mathrm{perm}} \geq D_{\mathrm{obs}}$ and $B = 100{,}000$. The addition of 1 to both the numerator and denominator accounts for the observed labeling as one of the possible outcomes and prevents a Monte Carlo $p$-value of zero when no sampled assignment exceeds the observed statistic.[51]

**Bootstrap confidence intervals for spatial speed profiles.** For each ROI and selected time point, PIV-derived speed was averaged along the vertical direction at each horizontal position, yielding one complete speed-versus-distance profile per ROI (Supplementary Fig. 9a). For each experimental condition and shear state, the mean spatial profile was calculated by averaging the ROI-level profiles, giving equal weight to each ROI. Uncertainty in the mean profile was estimated using non-parametric bootstrap resampling of complete ROI profiles with replacement. In each bootstrap replicate, the same number of ROI profiles as in the original group was sampled, and the mean profile was recalculated at every horizontal position. This procedure was repeated 5,000 times. The shaded bands in Supplementary Fig. 9 represent pointwise 95% percentile-bootstrap confidence intervals, defined by the 2.5th and 97.5th percentiles of the bootstrap distribution of the mean at each horizontal position. Complete ROI profiles, rather than individual PIV grid points, were resampled, thereby preserving the spatial structure within each ROI and preventing the number of spatially correlated grid measurements from determining the confidence-interval width.

**Strain, MSD exponents, and cell shape comparisons**. These statistical comparisons were performed using the Mann–Whitney U test, also known as the Wilcoxon rank-sum test, as specified in the corresponding figure captions and Supplementary Table 1. This non-parametric test was used for comparisons of independent samples when the data were not assumed to be normally distributed. The test was performed using the built-in *ranksum* function in MATLAB. For the shear-strain and rheological analyses, each independent disc pull contributed one strain or fitted β value. For the MSD analysis, each tracked cell contributed one fitted exponent, α. Pooled PIV-derived speed measurements were used only to visualize the speed distributions and were not subjected to inferential statistical testing.

## Cell Cycle Analysis

To investigate the effect of mechanical perturbation on cell cycle progression, we performed time-lapse imaging following shear experiments, with FUCCI-expressing MDCK cells[19], under the Unconfined (UC) condition. The FUCCI (Fluorescent Ubiquitination-based Cell Cycle Indicator) system enables real-time visualization of cell cycle dynamics in living cells through a dual-color reporter: nuclei appear red in the G1 phase and transition to green during S/G2/M phases[23]. This approach allows spatiotemporal monitoring of cell cycle progression in individual cells during tissue deformation.

Time-lapse imaging was conducted under the same environmental and optical conditions described previously, capturing cell behavior over several hours following the mechanical perturbation. To analyze cell cycle progression, we used ConfluentFUCCI[24], an open-source GUI-based framework specifically developed for automated analysis of highly confluent and dynamic epithelial monolayers. These outputs were post-processed using a custom MATLAB script to identify the time points (frames) at which individual cells transitioned from red to green/yellow, indicating progression from G1 to S phase. This enabled quantitative assessment of how many cells advanced through the cell cycle during the experiment, and whether this progression was affected by the applied mechanical stress.

## Magnetic Tweezer

The magnetic tweezer was fabricated from a 100 mm long high-permeability ST37 low-carbon steel core, comprising a 70 mm cylindrical section and a CNC-sharpened conical tip with a 16° angle. The core was wound with 350 turns of 0.50-mm insulated copper wire, which was subsequently covered with insulating tape.

Magnetic force was calculated by pulling individual discs through PDMS and converting their measured kinematics into forces using a force balance. Discs were positioned 1 mm from a horizontally mounted magnetic tweezer and imaged while being pulled toward the tweezer tip. Disc trajectories were extracted using FIJI and MATLAB, and forces were computed from the measured velocities using a constant drag coefficient. The drag coefficient was independently determined by analyzing the sinking motion of identical discs in PDMS at 20 °C. Disc kinematics were extracted using FIJI and MATLAB, and assuming force balance at steady state and a disc mass of 0.2 mg. Measurements from 26 discs yielded an average drag coefficient of $C_D = 5 \times 10^6$. The resulting force–distance relationships were fitted to exponential functions and corrected for the 45° tweezer angle used in tissue experiments (Supplementary Fig. 2).

### MSD calculation

Mean squared displacement (MSD) analysis was performed to quantify spatial patterns of cell motility across experimental groups. Cell trajectories were obtained from TrackMate outputs, which provided the (x,y) centroid positions of individual cells in each frame. For each trajectory, MSD values were computed for increasing time lags $\Delta t$ according to the standard definition:

$$MSD(\Delta t) = \langle [x(t+\Delta t) - x(t)]^2 + [y(t+\Delta t) - y(t)]^2 \rangle_t$$

where t denotes averaging over time. MSD was calculated up to a maximum lag of 180 minutes. Tracks shorter than 5 frames were excluded from the analysis. For each condition, MSD curves were computed for all individual tracks.

To characterize the mode of transport, the MSD curve of each individual track was fitted to a power-law model:

$$MSD(\Delta t) \propto (\Delta t)^{\alpha}$$

The exponent $\alpha$ serves as an indicator of the motion type. Due to the non-normal distribution of exponents across the population, the median exponent was utilized as the primary metric for central tendency. To estimate the uncertainty of these population medians, non-parametric bootstrap resampling was performed. For each experimental condition, the distribution of $\alpha$ values were resampled 1,000 times with replacement to generate a bootstrap distribution of the median using MATLAB ***bootci*** function[52]. The error bars are the resulting 95% confidence intervals (CI) of the median.

Because of substantial trajectory-to-trajectory variability, α was interpreted as a population-level descriptor rather than as a characteristic individual-trajectory phenotype.

## Acknowledgements

L.D acknowledges financial support from the Israel Scholarship Education Foundation (ISEF). L.A acknowledge financial support by the Israeli Science Foundation (ISF) on the individual research grant number 2107/21, and the New-Faculty Equipment grant number 2108/21; Israeli Ministry of Innovation, Science and Technology 0357- 24; BGU's joint collaboration grant, Engineering Sciences and the Blaustein Institutes for Desert Research, 2025; Pearlston Center at BGU; Israeli Young Academy.

## Data sharing

Data will be shared upon reasonable request from the corresponding author.

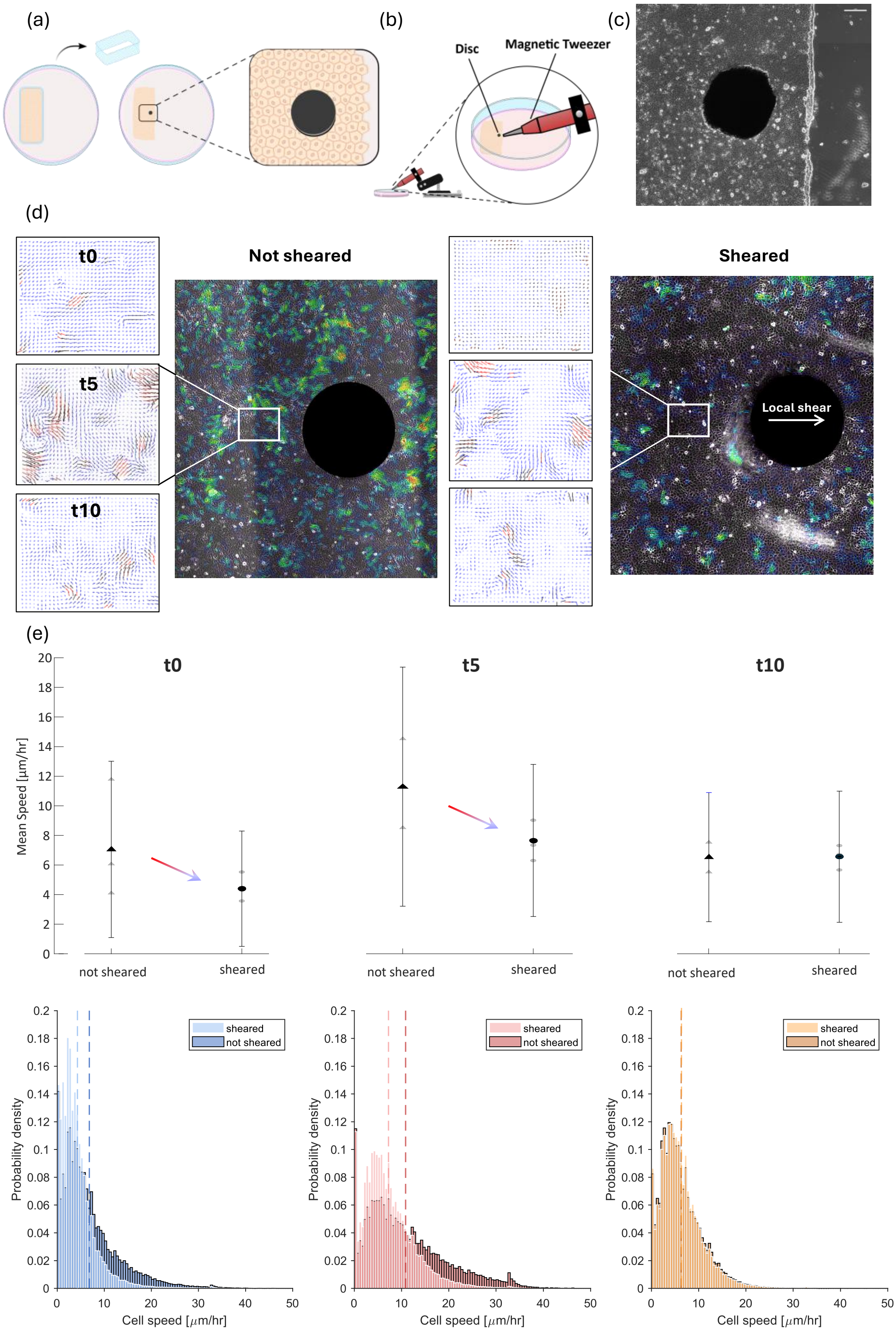

**Figure 1. Local mechanical shear suppresses cellular dynamics many cell rows away from the site of perturbation.** (a) Schematic of the epithelial experimental assay using MDCK cells cultured in ibidi wells (Methods). (b) Illustration of the experimental setup showing the magnetic tweezer and the attached magnetic disc (750 µm diameter, 500 µm height; Supplementary Fig. 1) used to apply localized shear stress (see Methods for force calibration; Supplementary Fig. 2). (c) Phase-contrast image of the epithelial monolayer with the magnetic disc in place (scale bar, 200 µm). (d) Representative PIV maps for sheared and not-sheared conditions (disc region was masked to reduce noise). The white rectangle marks the region of interest (ROI, 402 × 332 µm) used for analysis.

The right PIV maps are shown after 10s of magnetic force application for each maturation time, where t0 is defined as the time at which cells became fully confluent. (t0 = 20 hr, t5 = 25 hr, t10 = 30 hr; scale bar, 100 µm; Supplementary Video 2). (e) Top, speeds across different maturation times (within the ibidi wells). Data were collected from three ROIs (the presented white rectangle, with two more identical ROIs above and below it; 29X35 PIV grid) per condition across multiple independent plates: t0 sheared (n=3), t0 not sheared (n=3), t5 sheared (n=3), t5 not sheared (n=2), t10 sheared (n=3), and t10 not sheared (n=2). Time-lapse imaging was performed over 3–5 hours at 10-minute intervals. Gray symbols represent the mean speed from each independent experiment after averaging across the three analyzed ROIs and the complete imaging period. Black symbols and error bars represent the mean and standard deviation across independent experiments. Bottom, within each maturation condition, all experiments were analyzed identically and contributed the same number of measurements to the pooled sheared and non-sheared distributions. Identical bin edges were used for the two groups. Dashed vertical lines indicate the corresponding pooled means. Because individual PIV grid values are spatially and temporally correlated, statistical comparisons of the complete speed distributions were performed using an ROI-label permutation test with a KS-distance statistic. This analysis showed a shear-associated shift toward lower speeds at t0 and t5 ($p = 0.0383$ and $p = 0.0252$, respectively), but not at t10 ($p = 0.9007$) (Methods; Supplementary Fig. 8).

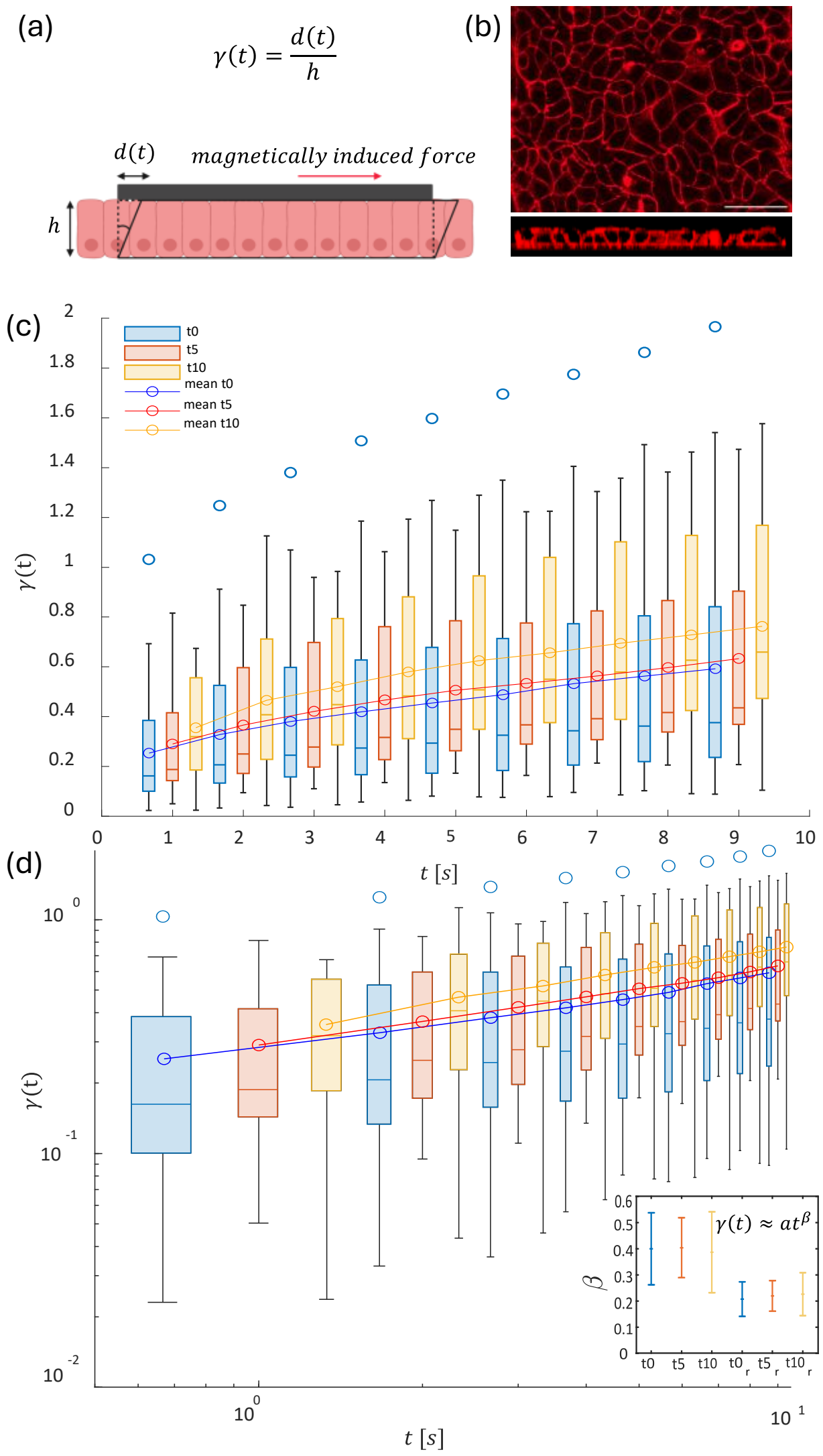

**Figure 2. Shear strain measurements reveal how, during maturation, the tissue softens.** (a) Schematic of the tissue shear model used to assess shear strain. (b) Fixed and stained epithelial monolayers used to assess the tissue height (scale bar, 50 µm; see Methods). (c) Boxplot of shear strain $\gamma(t) = \frac{disc\ displacement}{tissues\ height}$ distributions across maturation stages, obtained from independent experiments: t0 (n=30); t5 (n=14); t10 (n=27) (Supplementary Fig. 5). Boxplots display the median (horizontal line inside each box), the mean (circle), and the mean trend (connected line across means). Boxes indicate the interquartile range, while the whiskers extend to the most extreme data points within 1.5 times the interquartile range, and individual circles represent outliers beyond this range. (d) Boxplot of the log-log-transformed data. The inset shows β values extracted from power-law fits to the shear strain curves, shown for each maturation stage. r denotes recovery. Error bars represent standard deviation. Average $a$ values for each maturation stage: $t0$: $\bar{a} = 0.25$, $t5$: $\bar{a} = 0.28$, $t10$: $\bar{a} = 0.35$, $t0_r$: $\bar{a} = -0.21$, $t5_r$: $\bar{a} = -0.29$, $t10_r$: $\bar{a} = -0.27$. Statistical comparisons were performed using the Mann-Whitney rank-sum test (Supplementary Table 1).

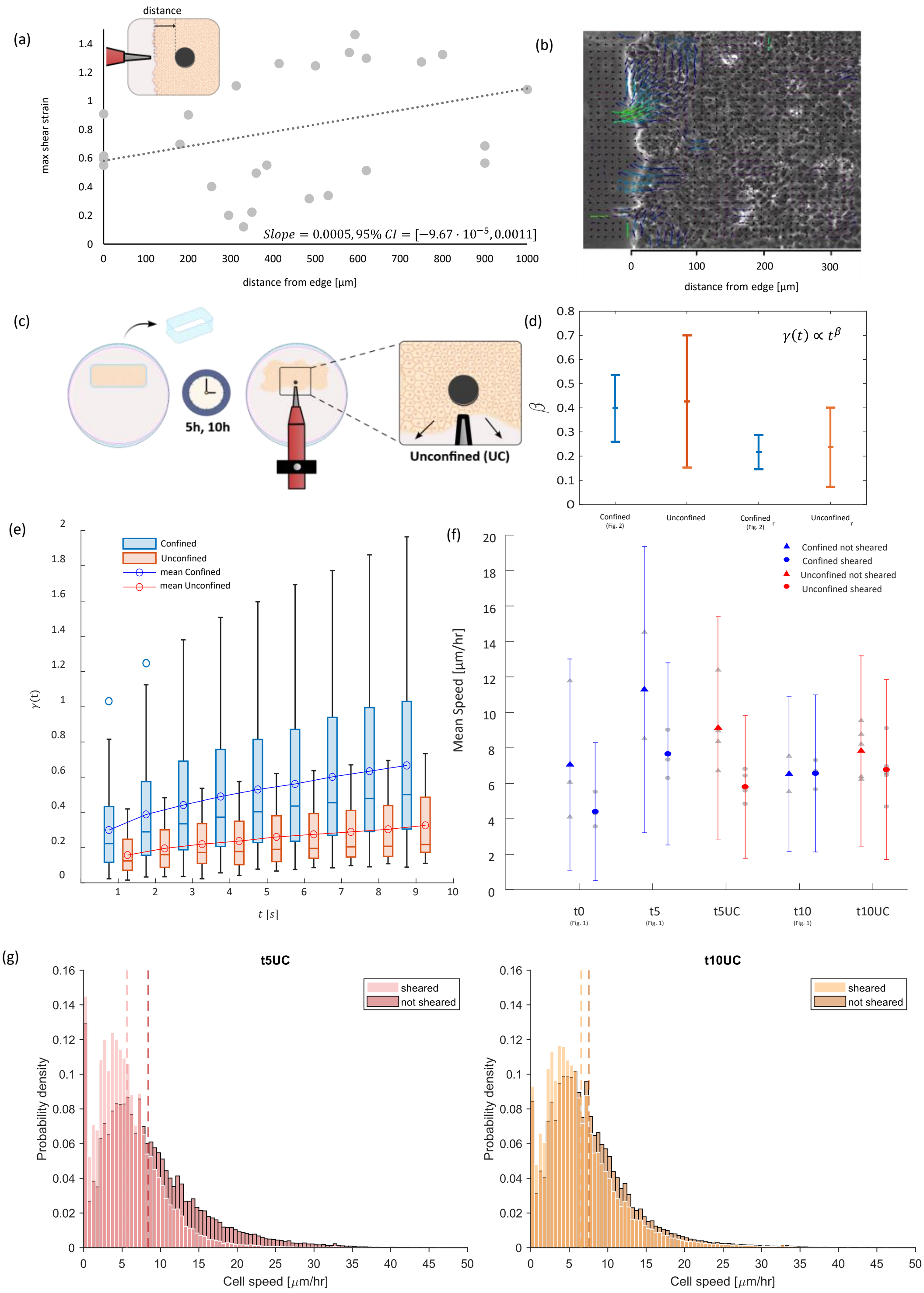

**Figure 3. Unconfinement increases tissue dynamics and stiffness, and enables long-range shear-induced suppression.** (a) The maximum shear strain $\gamma$ as a function of the disc's distance from the tissue edge. Data were obtained from confined t10 experiments (n = 27). Pearson correlation analysis $r = 0.399$. (b) PIV analysis at the tissue edge, demonstrating elevated cellular motility near the edge. (c) Schematic of the Unconfined (UC) experimental assay. Culture well is removed after 20 hours, upon reaching confluency. (d) $\beta$ values extracted from power-law fits of shear strain curves for each condition (Confined and Unconfined), where r denotes recovery. Error bars represent standard deviation. Average $a$ values for each condition: $confined{:}\bar{a} = 0.3, unconfined{:}\bar{a} = 0.26, confined_r{:}\ \bar{a} = -0.25, unconfined_r{:}\ \bar{a} = -0.15$. (e) Comparison of shear strain over time between Unconfined and Confined Independent experiments: t5 UC (n= 10); t10 UC (n= 13). Boxplots represent shear strain distributions for each condition, and display the median (horizontal line inside each box), the mean (circle), and the

mean trend (connecting line across means). Boxes indicate the interquartile range, while the whiskers extend to the most extreme data points within 1.5 times the interquartile range, and individual circles represent outliers beyond this range. (f) Mean speed over time for each condition; red represents Unconfined experiments, blue represents the Confined data as shown in Figure 1. UC Data were collected from 3 ROIs (29X35 Grid) per condition across multiple independent plates: t5 UC sheared (n=5), t5 UC not sheared (n=4), t10 UC sheared (n=5), and t10 UC not sheared (n=5). Gray symbols represent the mean from each independent experiment, and colored symbols and error bars represent the mean and standard deviation across independent experiments. (g) Distributions of local PIV-derived speeds in sheared and non-sheared unconfined monolayers at t5UC and t10UC. Within each unconfined maturation condition, each experiment contributed the same number of observations to the pooled sheared and non-sheared distributions. Identical bin edges were used for the two groups. Dashed vertical lines indicate the corresponding pooled means. The same ROI-level analysis as in Fig. 1 showed shear-associated shifts toward lower speeds under unconfined conditions at both t5UC and t10UC ($p = 1.60 \times 10^{-4}$ and $p = 0.0295$, respectively) (Methods, Supplementary Fig. 8).

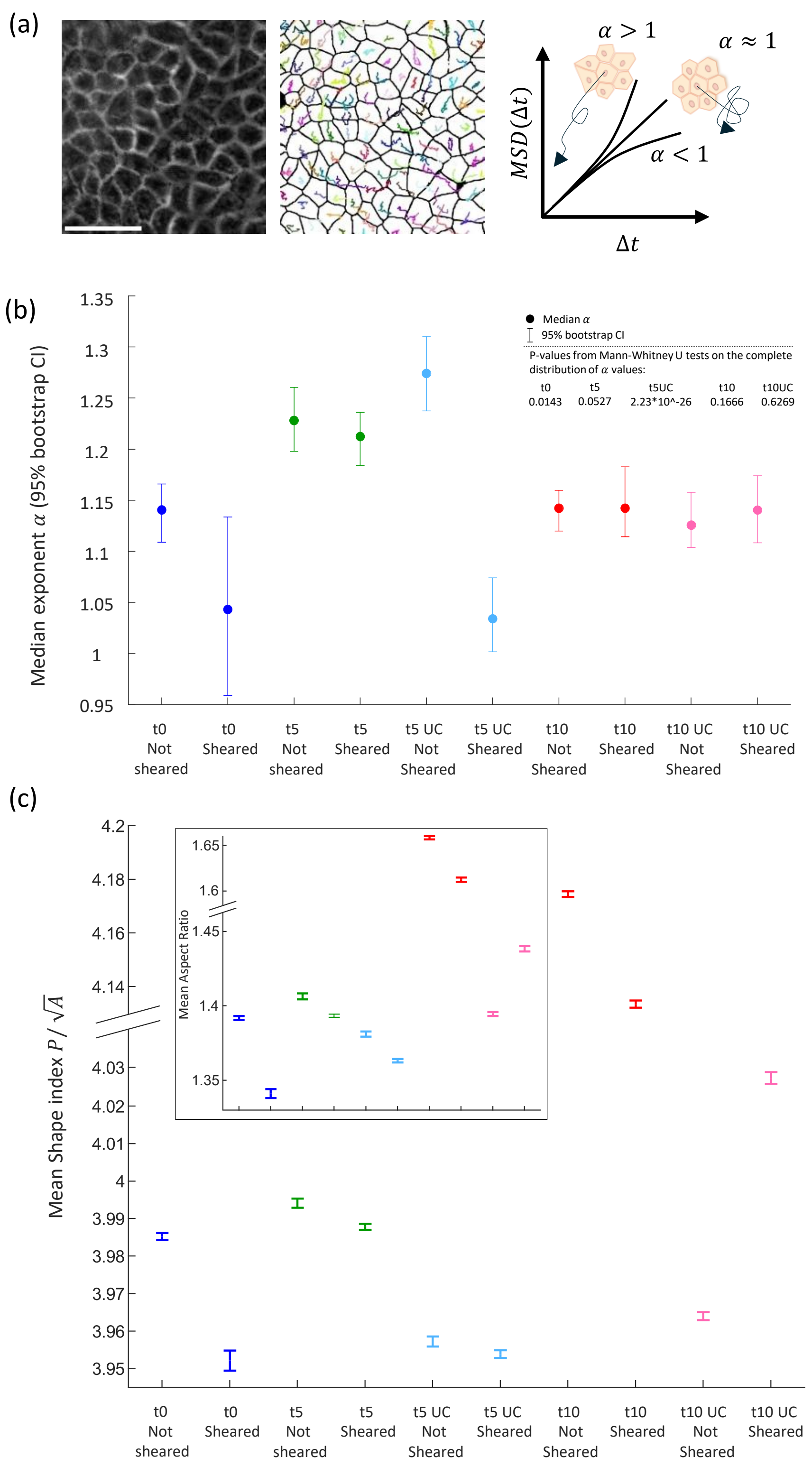

**Figure 4. Shear-induced migratory suppression partially coincides with epithelial jamming metrics.** (a) Phase-contrast image of the region of interest (scale bar, 100 µm) used for velocity analysis (left), corresponding cell segmentation generated using Cellpose (see Methods), and individual cell tracks (see Methods; middle) and illustrating caging dynamics as portrayed by exponent $\alpha$ (right). (b) Median of the $\alpha$ values obtained from the power-law model ($MSD\,(\Delta t) = At^{\alpha}$) fit for each track. Error bars denote the 95% bootstrap confidence intervals (CIs) around the median, calculated using 1,000 bootstrap resamples per condition (methods). Outliers of negative $\alpha$, and $\alpha > 2$ values were filtered out. Time-lapse imaging was performed for 3 hours at 10-minute intervals. t0 sheared (n=3, 4 ROIs in total with 406 Average number of cells); t0 not sheared (n=3, 9 ROIs with 2077 Avg. ); t5 sheared

(n=4, 12 ROIs with 3051 Avg.); t5 not sheared (n=1, 3 ROIs with 1208 Avg.); t5 UC sheared (n=5, 13ROI total with 1812 Avg); t5 UC not sheared (n=2, 5ROI total with 1453 Avg.); t10 sheared (n=1, 3ROI with 1785 Avg.); t10 not sheared (n=2; 6ROI total with 3230 Avg.); t10 UC sheared (n=3, 7ROI total with 1650 Avg.); t10 UC not sheared (n=4, 11ROI total with 2242 Avg). Number of MSD curves for each condition: t0 not sheared (n=2379); t0 sheared (n=407); t5 not sheared (n=1547); t5 sheared (n=3918); t5 UC not sheared (n=1916); t5 UC sheared (n=2344); t10 not sheared (n=3619); t10 sheared (n=1934); t10 UC not sheared (n=3319); t10 UC sheared (n=2454). The inset reports p-values from pairwise Mann–Whitney rank-sum tests comparing the complete distributions of α values between sheared and not-sheared conditions. (c) Mean cell shape ($Cell\ Perimeter/\sqrt{Cell\ Area}$) and aspect ratio ($Long\ major\ axis/short\ major\ axis$) for each experimental condition. Calculations are based on the same segmented cell data used for the MSD analysis. Symbols represent the mean across all segmented cells in each condition, and error bars denote the standard error of the mean (SEM).

# Supplementary Information

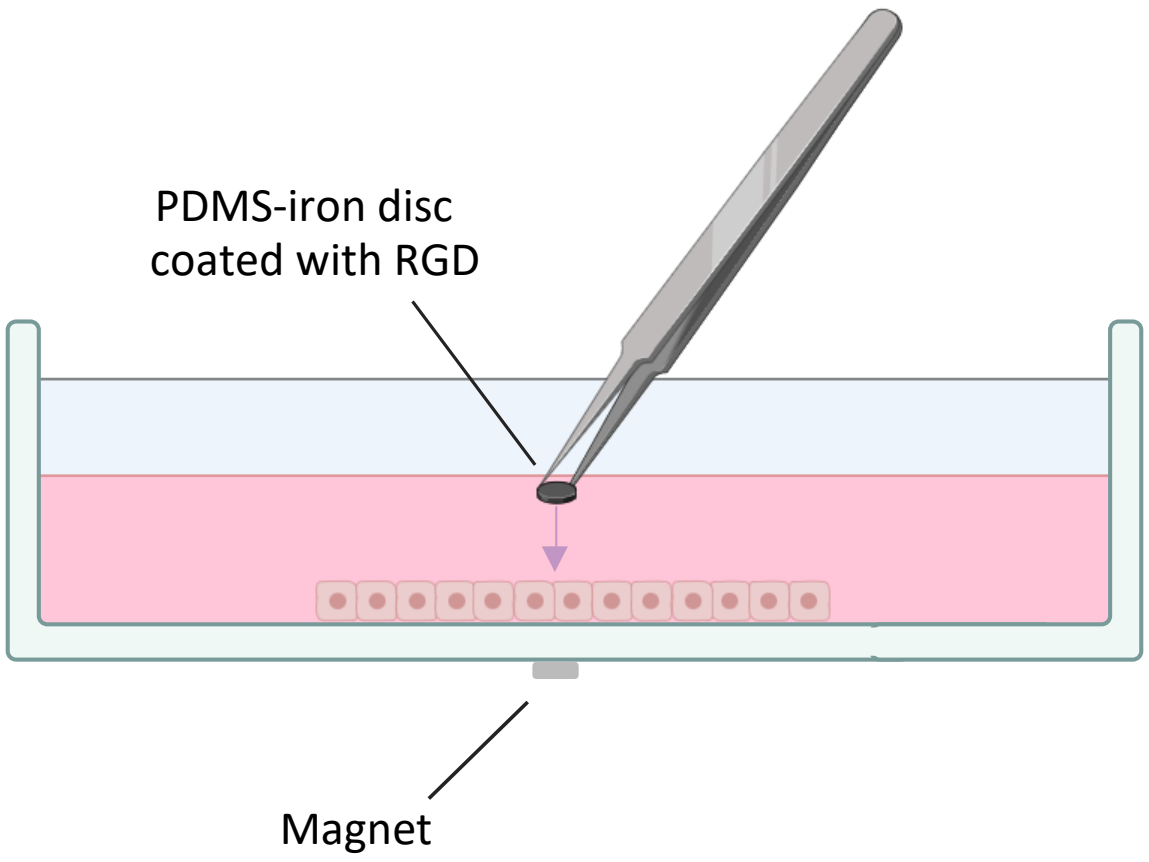

**Supplementary Figure 1. Magnetic disc placements.** Illustration of the method used to position the magnetic disc on the tissue, using tweezers and a magnet below the dish (Methods). The disc surface was activated with Sanpah and coated with RGD peptide that adheres to the cells.

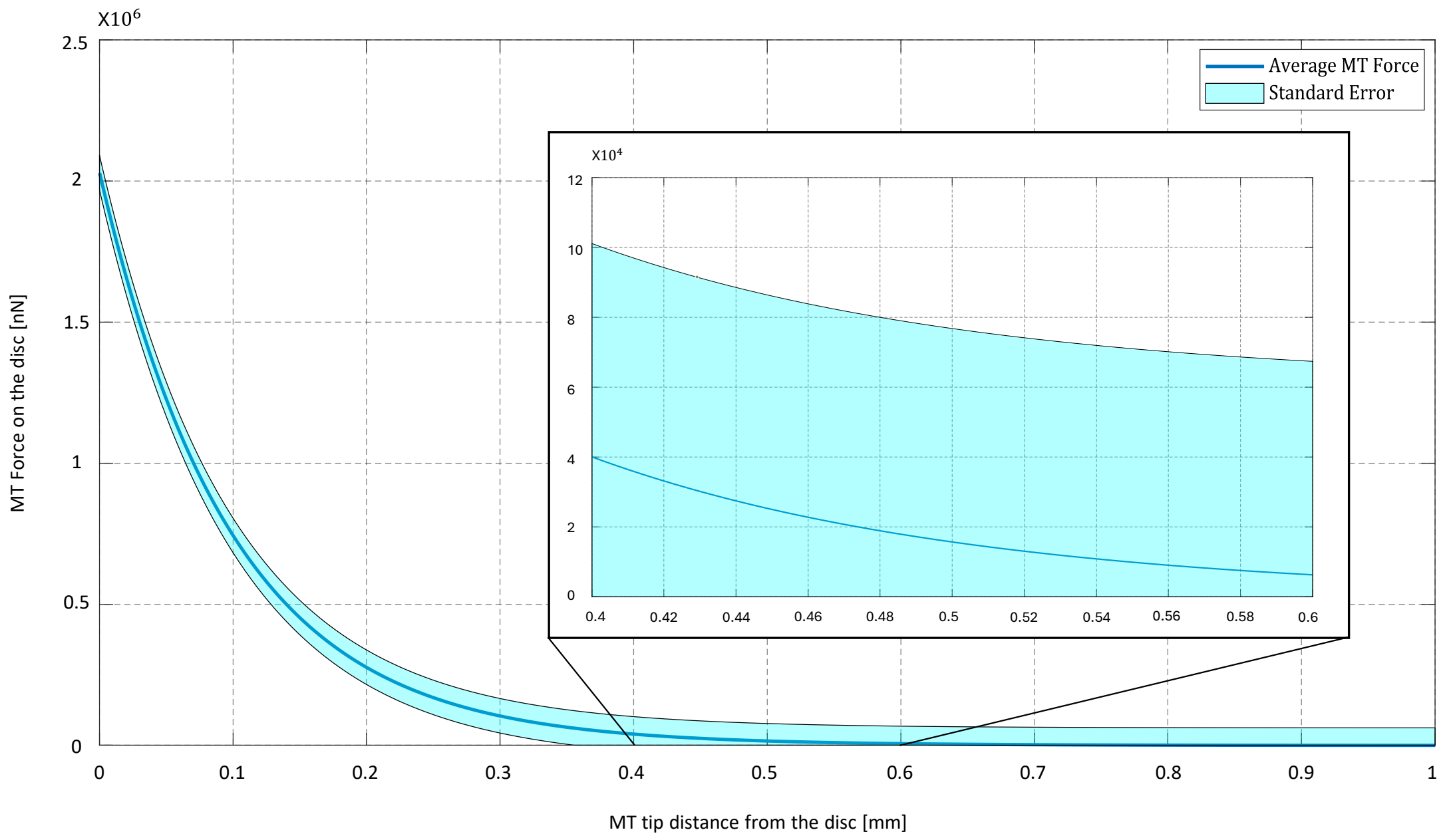

**Supplementary Figure 2. Force calibration.** Average force exerted on the magnetic disc in MDCK shear experiments as a function of distance from the magnetic tweezer (MT) over the full 0–1 mm range. Inset: The average force within the 0.4–0.6 mm range. 500 µm distance between the MT and the magnetic disc produces a force of $\sim 2 \cdot 10^4$ [nN].

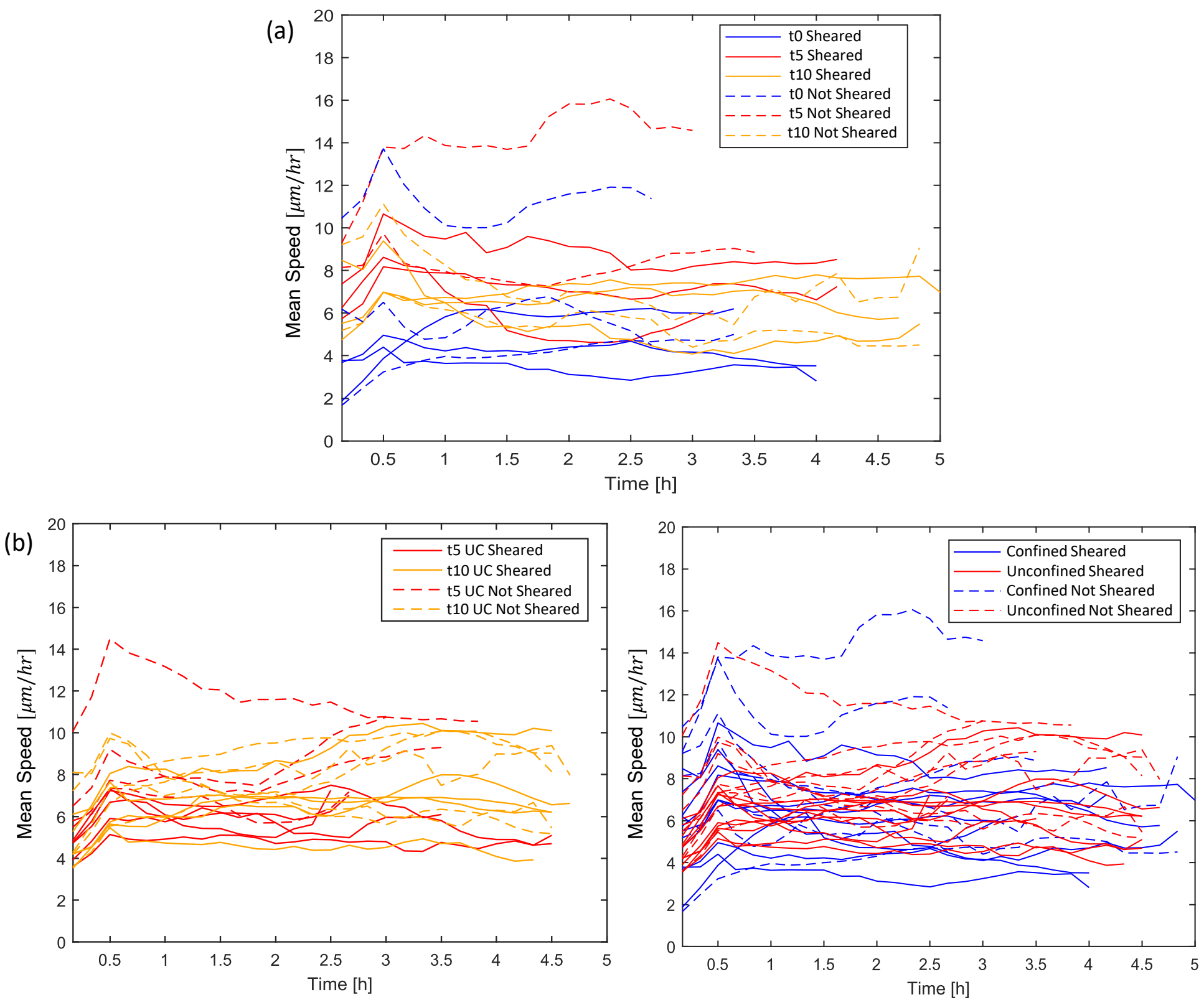

**Supplementary Figure 3. Mean speed over time across maturation times.** (a) Mean speed curves over time for different maturation times. Each curve represents an individual experiment. (b) (Left) Mean speed curves over time for different maturation stages in the Unconfined experiments. (Right) Comparison of mean speed over time between Confined and Unconfined experiments.

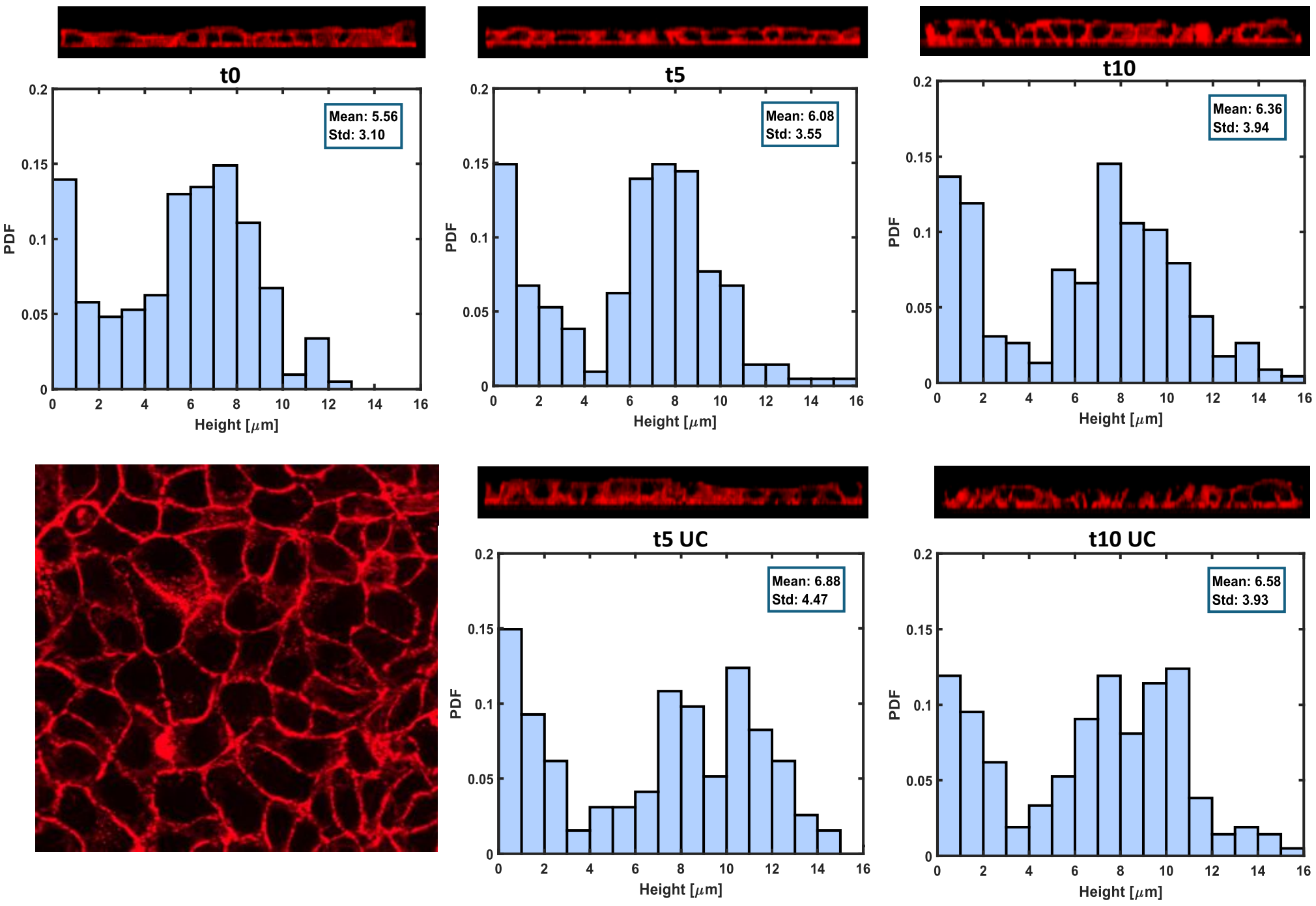

**Supplementary Figure 4. Tissue height measurements.** Fixed and stained epithelial monolayers were used to measure cell height across experimental conditions (t0, t5, t5UC, t10, t10UC). Probability density functions (PDFs) of cell height are shown for each condition, along with corresponding mean and standard deviation values.

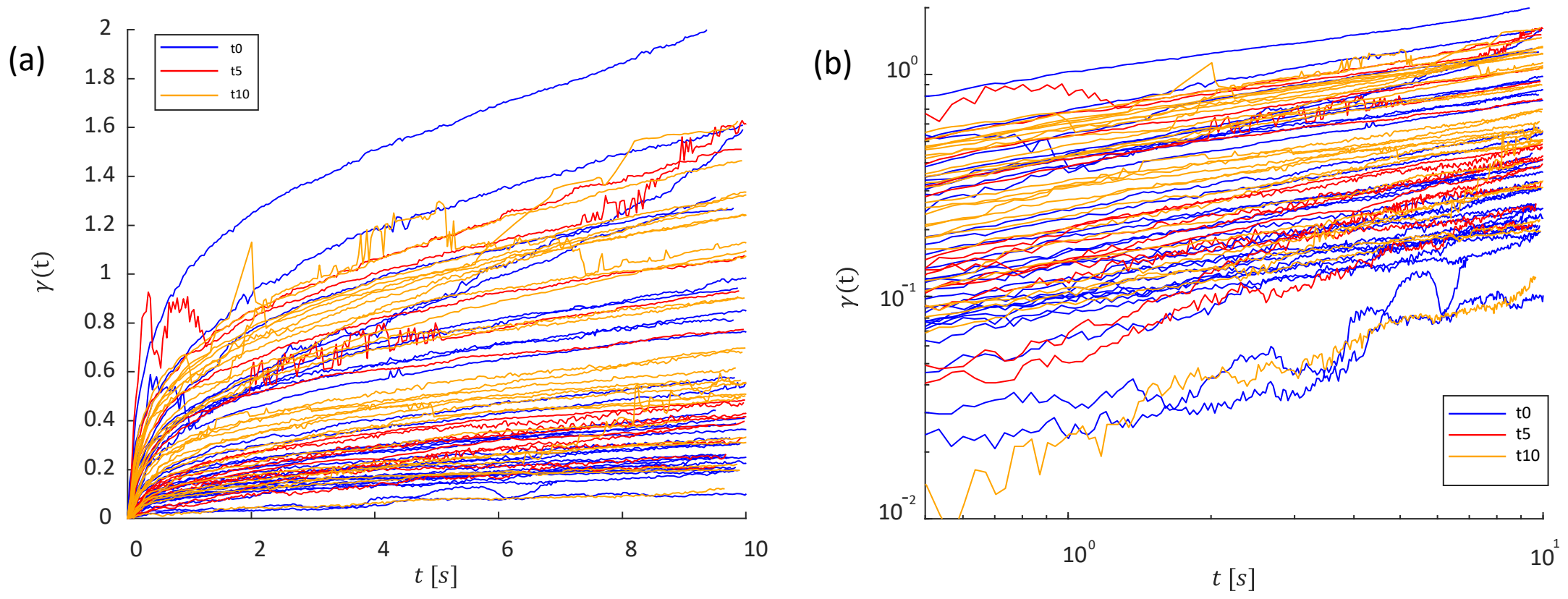

**Supplementary Figure 5.** Shear strain curves from the independent experiments, divided to maturation groups: t0 (n = 30), t5 (n = 14), and t10 (n = 27). Panels show (a) regular scale, (b) log–log scale.

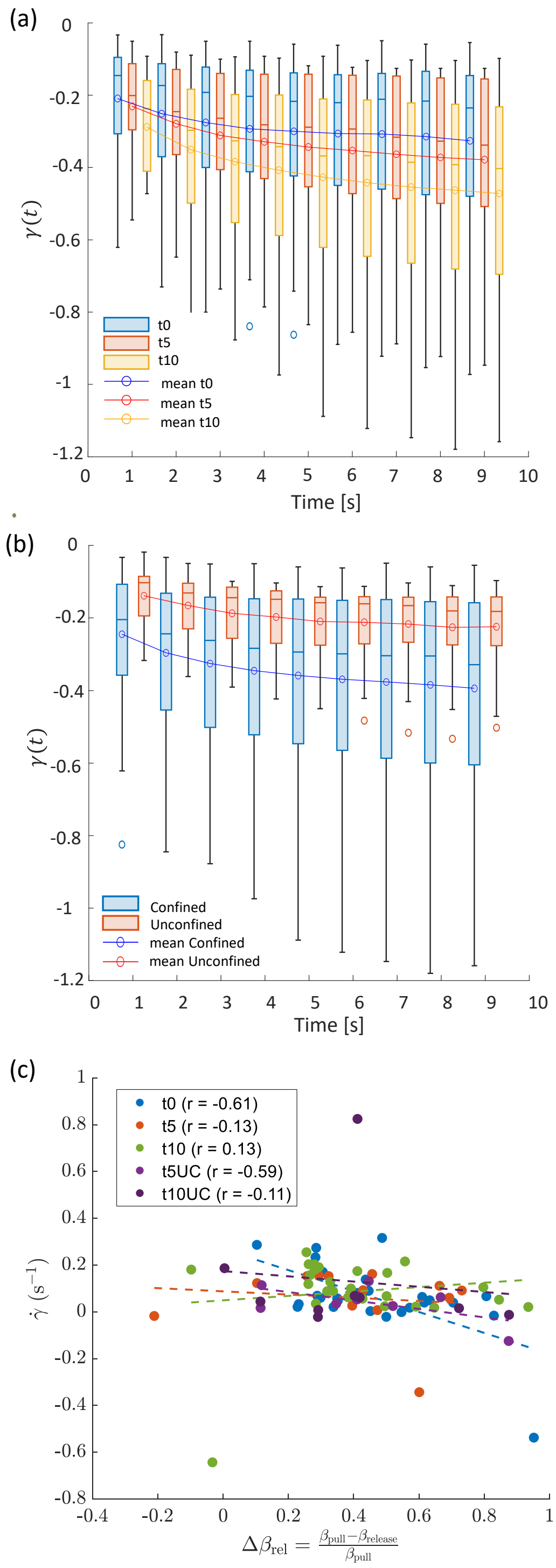

**Supplementary Figure 6. Release-phase shear-strain dynamics and the relationship between initial strain rate and the loading-to-release change in $\boldsymbol{\beta}$.** Boxplot shear strain curves of the release phase (a) divided by maturation groups: t0 (n = 30), t5 (n = 14), and t10 (n = 27). (b) divided by the different growth conditions: Confined (n=71), and unconfined (n=23). Boxplot elements are defined as in Fig. 2c. (c) Relationship between the initial strain rate, defined as the rate at 1s, and the relative change in the power-law exponent between the loading and release phases for individual disc pulls. The relative change was defined as $\Delta\beta_{\rm rel} = (\beta_{\rm pull} - \beta_{\rm release})/\beta_{\rm pull}$. Because strain rate was obtained by numerical differentiation of the strain trajectory (methods), the estimate is sensitive to frame-to-frame displacement noise; occasional negative values may therefore occur. Dashed lines show separate linear regressions for t0, t5, t10, t5UC, and t10UC, and the corresponding Pearson correlation coefficients are reported in the legend. No consistent association between the initial strain rate and $\Delta\beta_{\rm rel}$ was observed across experimental conditions.

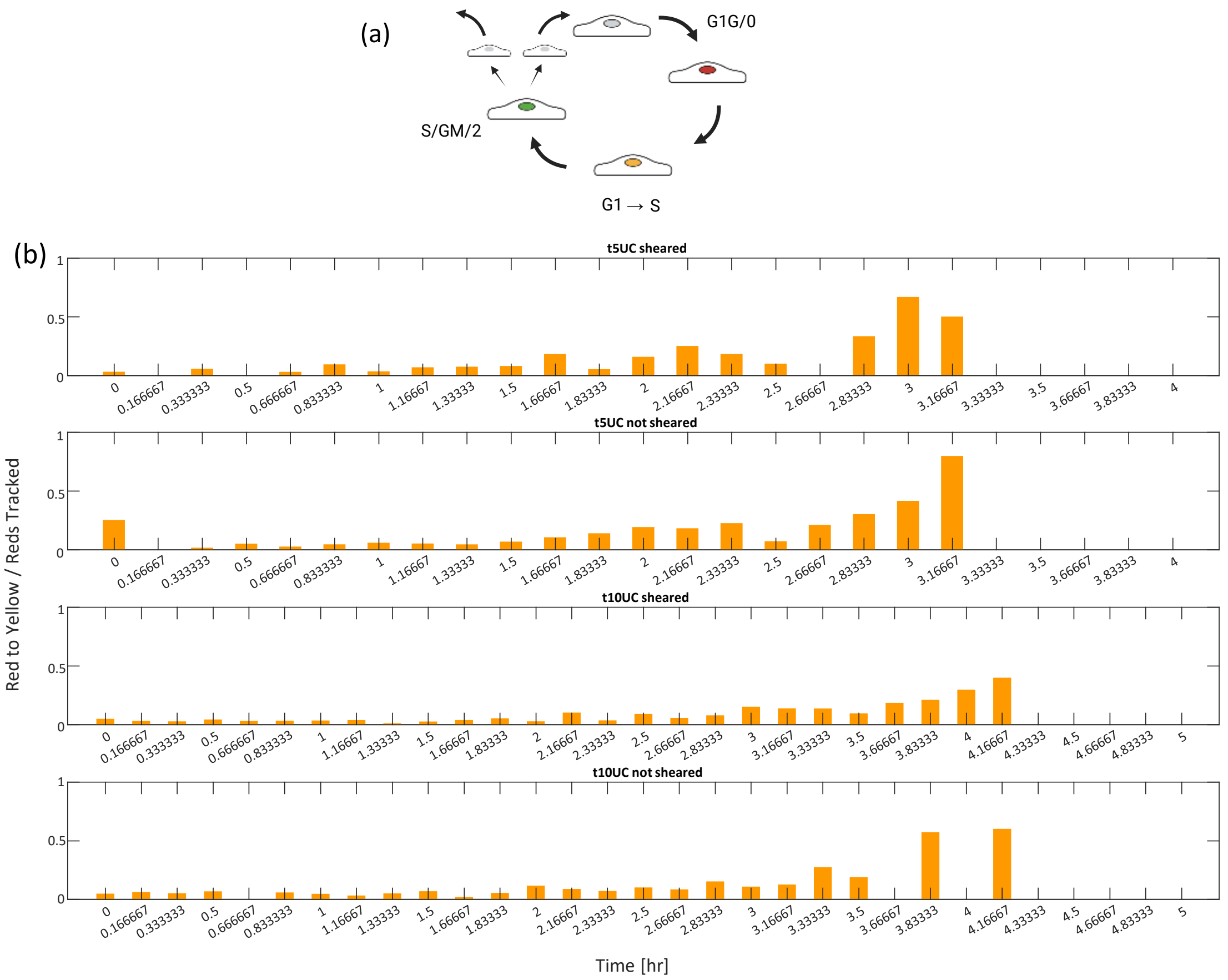

**Supplementary Figure 7. FUCCI cell cycle progression analysis for Unconfined experiments.** (a) Cell-Cycle illustration. (b) The fraction of cells transitioning to S-phase (yellow/green) divided by G1-phase (red) cells is shown for each UC condition: t5, t10, t5 not sheared, and t10 not sheared. Transitions were counted at the last frame in which the cell appeared red.

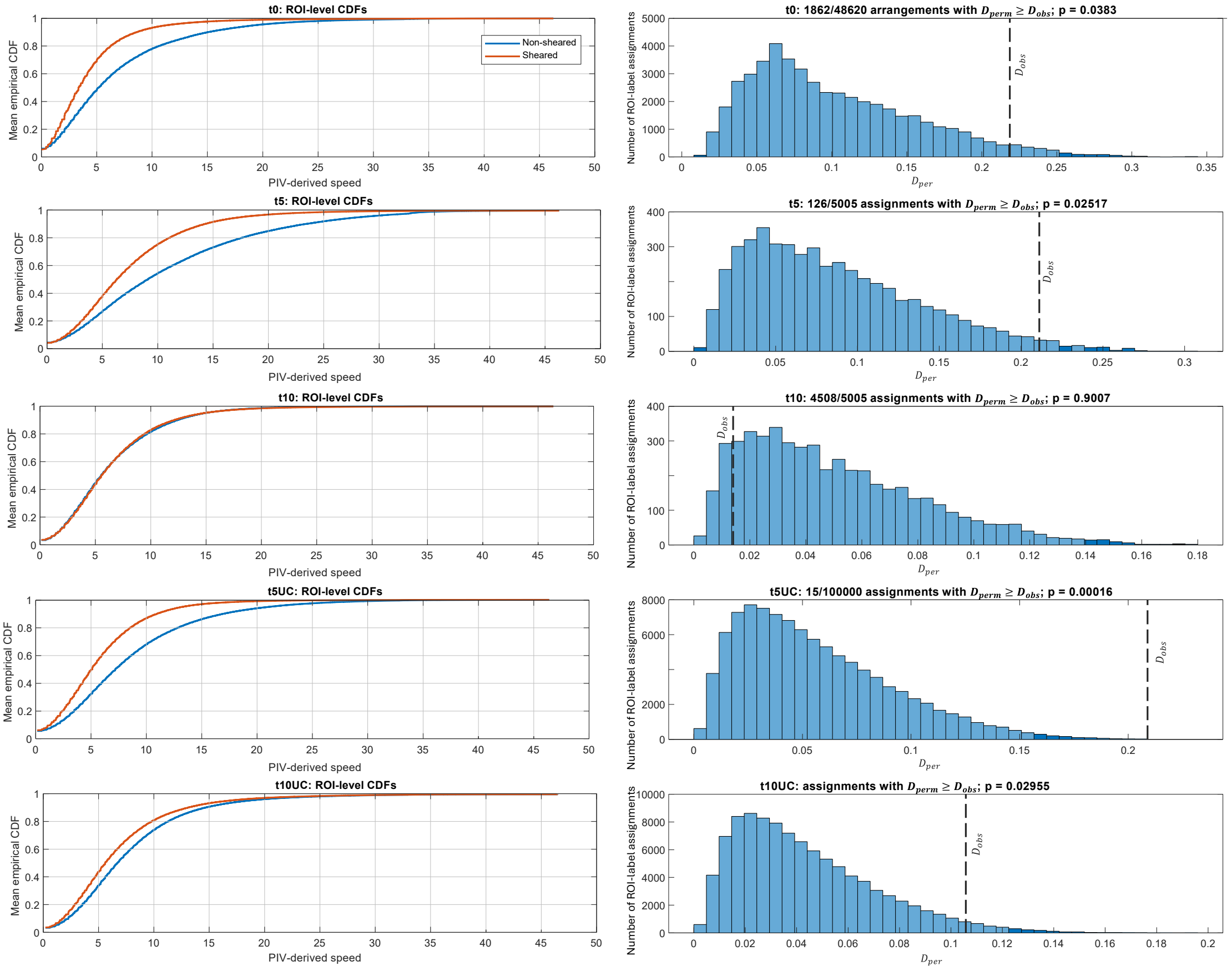

**Supplementary Figure 8. ROI-label permutation analysis of PIV-derived speed distributions using a KS-distance statistic.** Rows correspond to t0, t5, t10, t5UC, and t10UC, from top to bottom. Left, mean empirical cumulative distribution functions (CDFs) of PIV-derived speed for non-sheared (blue) and sheared (orange) ROIs. A normalized CDF was calculated for each ROI and then averaged within each group, giving equal weight to every ROI. Right, null distributions of the KS-distance statistic, $D_{\text{perm}}$, obtained by reassigning the sheared and non-sheared labels among ROIs while preserving the observed group sizes (methods). Dashed vertical lines indicate the observed KS distance, $D_{\text{obs}}$. The numbers of ROIs were 9 non-sheared and 9 sheared for t0, 6 and 9 for t5, 6 and 9 for t10, 12 and 15 for t5UC, and 15 and 15 for t10UC, respectively. All possible ROI-label assignments were evaluated for t0, t5, and t10, whereas 100,000 assignments were sampled for t5UC and t10UC. The corresponding $p$-values are shown above the right-hand panels. See Methods for full details.

| **Shear Strain comparison** | Condition | 1s | 2s | 3s | 4s | 5s | 6s | 7s | 8s | 9s |
|---|---|---|---|---|---|---|---|---|---|---|
| Fig. 2 | t0 vs t5 | 0.44217 | 0.35772 | 0.37105 | 0.27304 | 0.22168 | 0.2957 | 0.31836 | 0.34415 | 0.35753 |
| | t0 vs t10 | 0.02182 | 0.02798 | 0.03159 | 0.02473 | 0.02182 | 0.02578 | 0.05934 | 0.06628 | 0.06158 |
| | t5 vs t10 | 0.24263 | 0.30255 | 0.30255 | 0.27749 | 0.28984 | 0.26551 | 0.30255 | 0.26551 | 0.22117 |
| Fig. 3 | t5UC vs t10UC | 0.9737 | 0.9737 | 0.92121 | 0.92121 | 0.92121 | 1 | 0.86907 | 1 | 0.86026 |
| | C vs UC | 0.00369 | 0.00235 | 0.00115 | 0.00071 | 0.00060 | 0.00056 | 0.00036 | 0.00025 | 0.00046 |

| **$\beta$ comparison** | Shear | | Recovery | | Shear vs Recovery | |
|---|---|---|---|---|---|---|
| Fig. 2 | t0 vs t5 | 0.75278 | t0 vs t5 | 0.73629 | t0 | 8.89*10^-11 |
| | t0 vs t10 | 0.31017 | t0 vs t10 | 0.49319 | t5 | 1.39*10^-5 |
| | t5 vs t10 | 0.30255 | t5 vs t10 | 0.84458 | t10 | 1.52*10^-7 |
| Fig. 3 | C vs UC | 0.50639 | C vs UC | 0.89007 | C | 3.94*10^-21 |
| | | | | | UC | 0.00027 |

| **values from MSD $\alpha$ fit** | Sheared vs Not Sheared | t0 | t5 | t5UC | t10 | t10UC |
|---|---|---|---|---|---|---|
| Fig. 4 | | 0.0143 | 0.0527 | 2.23*10^-26 | 0.1666 | 0.6269 |

**Supplementary Table 1. Statistical comparison results**. Mann–Whitney rank-sum test $p$-values for shear-strain, $\beta$, and MSD-exponent $\alpha$ comparisons. PIV-derived speed-distribution comparisons were performed separately using the ROI-label permutation analysis described in Methods and Supplementary Figure 8.

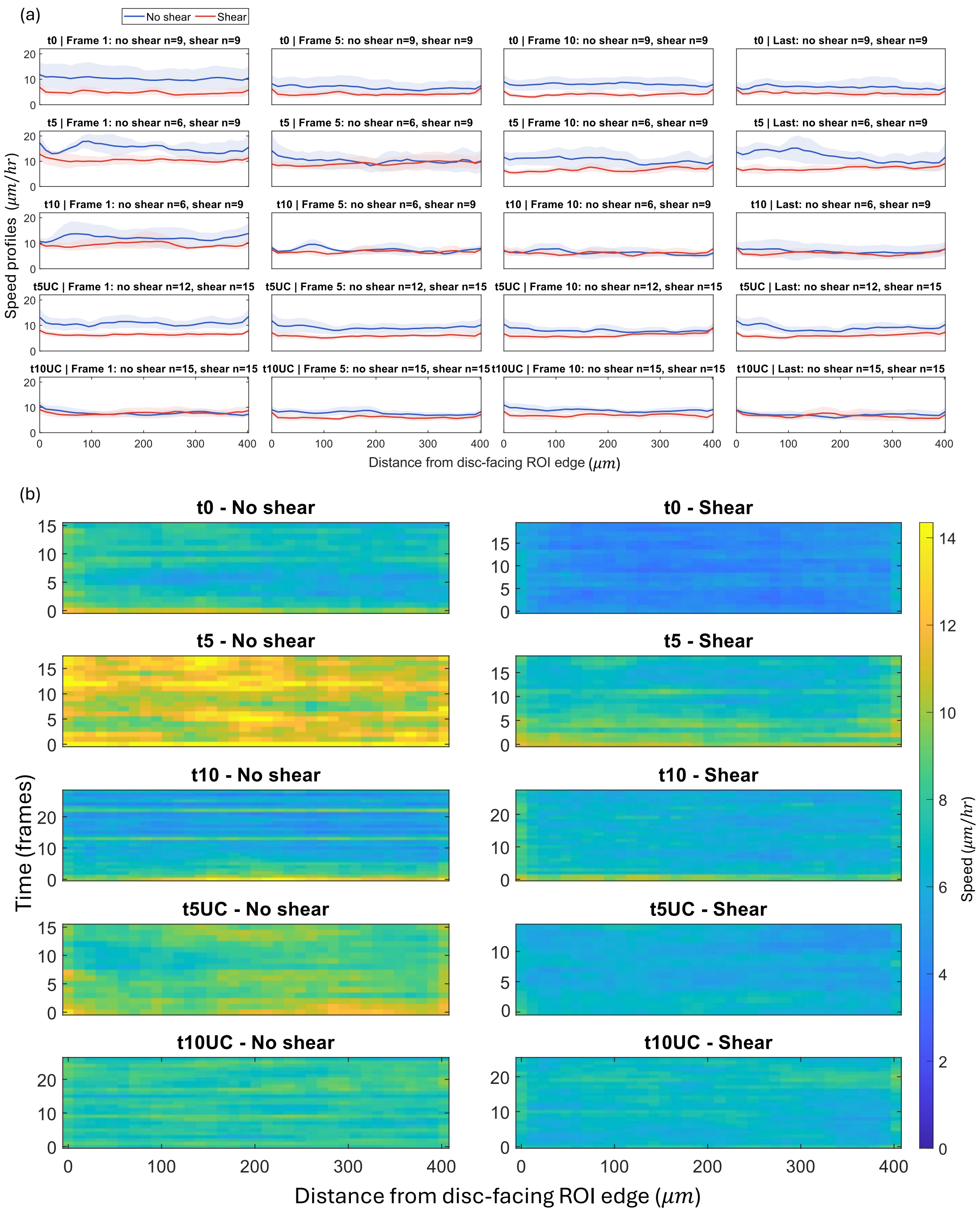

**Supplementary Figure 9. Speed profiles across the imaged field. (a)** cell speed as a function of distance from the disc-facing edge of the ROI at the first (10 min), fifth (50 min), tenth (100 min), and final retained time points (160 min for t0, 180 min for t5, 290 min for t10, 160 min for t5UC, and 270 min for t10UC). Rows correspond to t0, t5, t10, t5UC, and t10UC. Blue and red lines show the mean profiles across no-shear and shear ROIs, respectively; shaded regions indicate bootstrap 95% confidence intervals of the mean across ROI-level profiles (Methods). Speed profiles were obtained by averaging PIV-derived speed along the vertical direction of each ROI. **(b)** Corresponding kymographs showing PIV-derived speed as a function of horizontal distance and time for no-shear and shear samples.

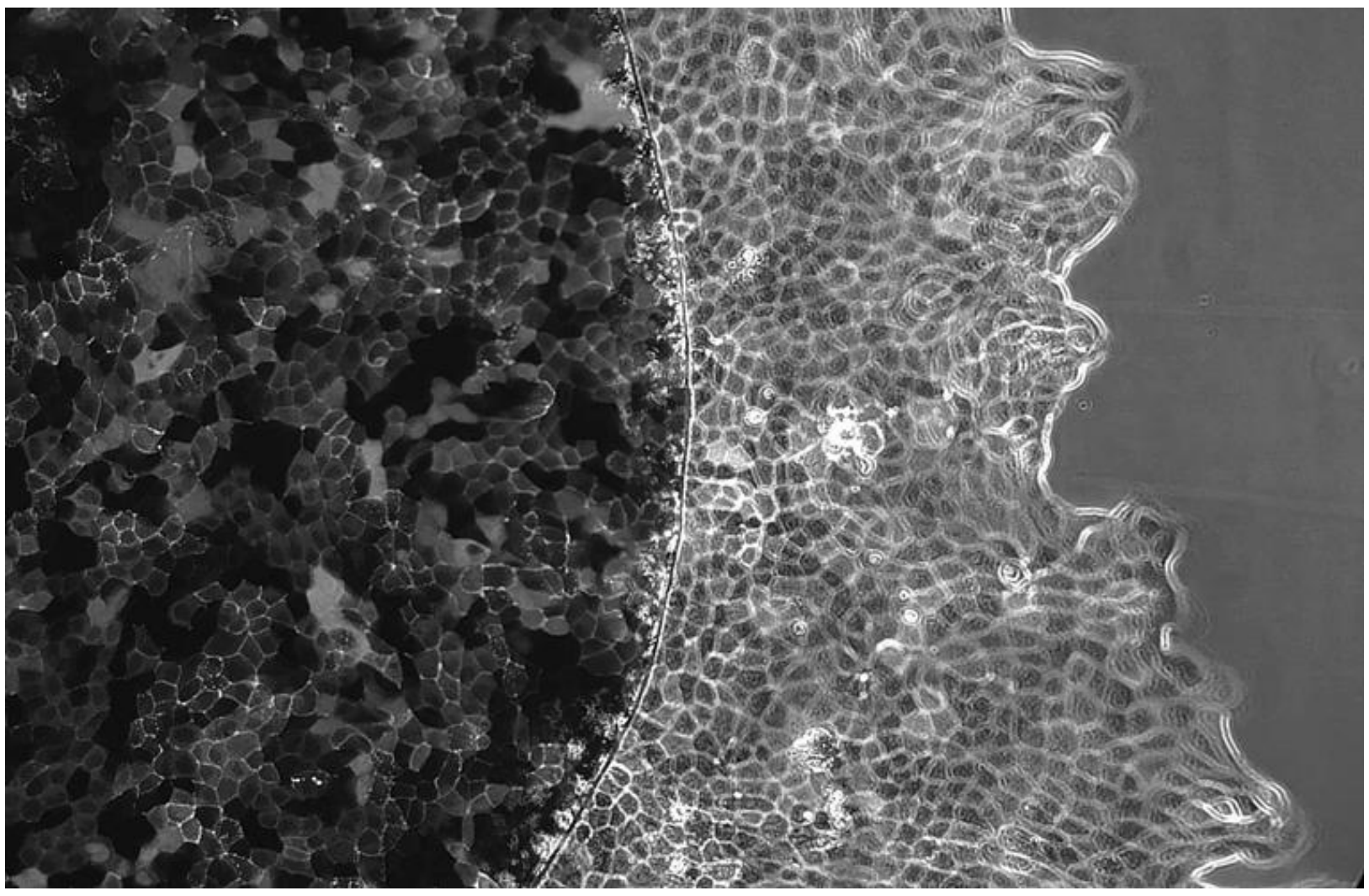

**Supplementary Video 1. magnetic shear and recovery.** A close-up of the magnetic disc and underlying cells during a 10-second force application followed by a 10-second recovery, visualized using ZO-1 GFP fluorescence markers.

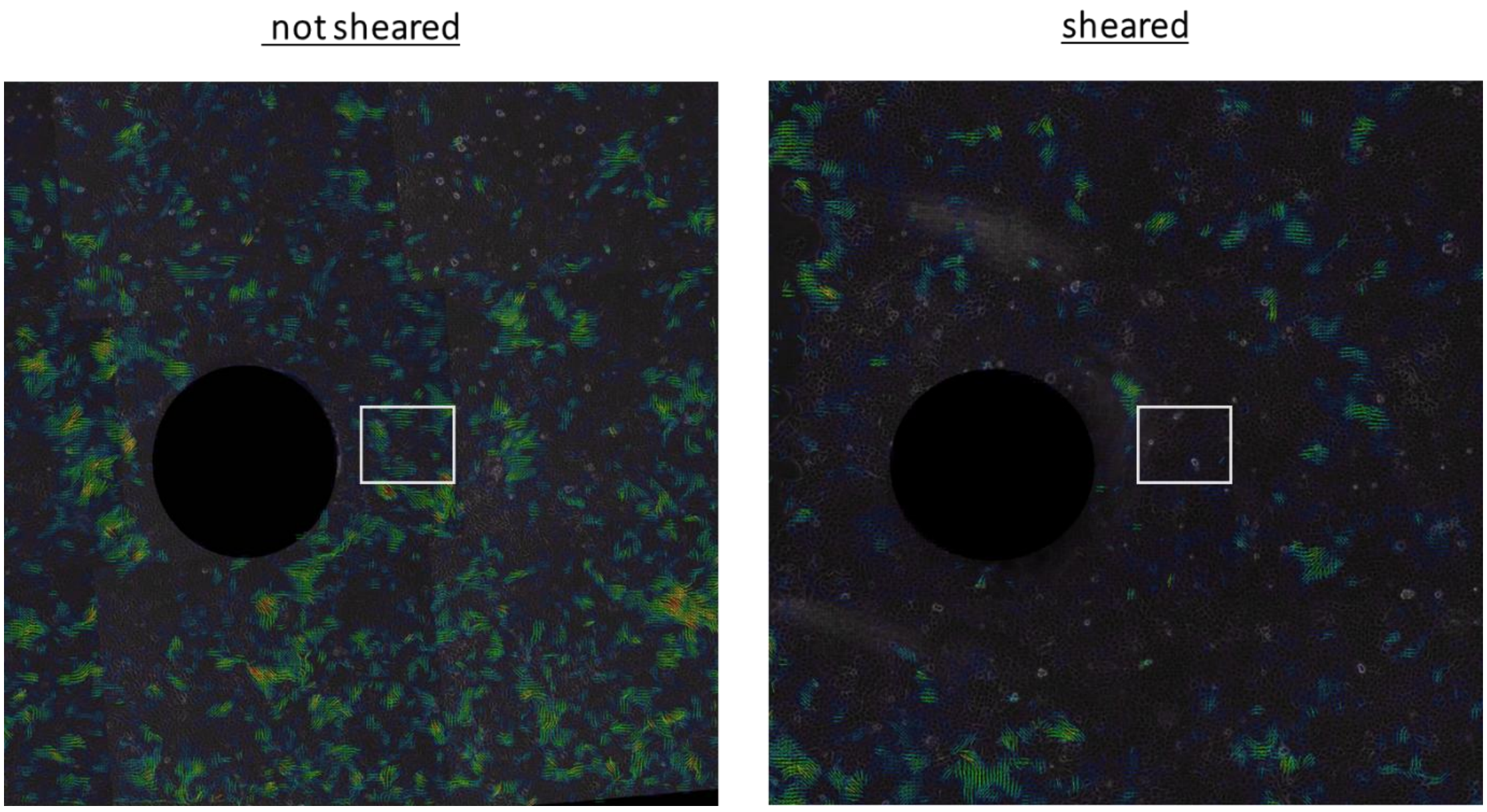

**Supplementary Video 2. Velocity fields around the magnetic disc in sheared and non-sheared tissues.** Representative PIV maps showing a wide region of the tissue surrounding the magnetic disc under sheared and non-sheared conditions. In the sheared condition, the 10-s shear perturbation was completed before the time-lapse sequence began and is not shown. The sequence begins approximately 10 min after shear, spans 3 hrs, and was acquired at 10-min intervals. The white rectangle indicates the region of interest used for quantifying mean speed.